\documentclass[11pt,twoside,a4paper]{article}
	\usepackage{graphicx}
	\usepackage{authblk}

\begin{document}

\title{A Prismatic Analyser concept for Neutron Spectrometers}
\author[a]{Jonas O Birk}{}{}
\author[b]{M\'arton Mark\'o}{}{}
\author[c]{Paul G Freeman}{}{}
\author[a]{Johan Jacobsen}{}{}
\author[d]{Niels B Christensen}{}{}
\author[b]{Christof Niedermayer}{}{}
\author[c]{Henrik M R\o nnow}{}{}
\author[a]{Kim Lefmann}{}{}

\affil[a]{Nano Science Center, Niels Bohr Institute, University of Copenhagen, DK-2100 Copenhagen \O, Denmark}
\affil[b]{Paul Scherrer Institute, Villigen, Scwitzerland}
\affil[c]{Laboratory for Quantum Magnetism, \'Ecole Polytechnique F\'ed\'erale de Lausanne (EPFL), 1015 Lausanne, Switzerland}
\affil[d]{Technical University of Denmark, DK-2800-Kgs. Lyngby, Denmark}
\maketitle


\begin{abstract}
A development in modern neutron spectroscopy is to avoid the need of large samples. We demonstrate how small samples together with the right choice of analyser and detector components makes distance collimation an important concept in crystal analyser spectrometers. We further show that this opens new possibilities where neutrons with different energies are reflected by the same analyser but counted in different detectors, thus improving both energy resolution and total count rate compared to conventional spectrometers. The technique can be combined with advanced focusing geometries and with multiplexing instrument designs. We present a combination of simulations and data with 3 energies from one analyser. The data was taken on a prototype installed at PSI, Switzerland, and shows excellent agreement with the predictions. Typical improvements will be 2 times finer resolution and a factor 1.9 in flux gain compared to a Rowland geometry or 3 times finer resolution and a factor 3.2 in flux gain compared to a single flat analyser slab.
\end{abstract}

\section{Introduction}
Most crystal analyser neutron spectrometers such as triple axis spectrometers rely on the analyser mosaicity to provide the desired compromise between intensity and energy resolution\cite{tripAxBook}. Coarser analyser mosaicity means reflection of a larger energy range resulting in higher recorded flux but coarser resolution, while fine mosaicity brings the opposite result. For cold neutron spectrometers the most common analyser material is Pyrolytic Graphite (PG), using the (002) reflection with typical mosaicity of 20' to 40' ($\frac{1}{3}^\circ-\frac{2}{3}^\circ$) FWHM as seen e.g. on TASP \cite{tasp}, PANDA \cite{PandaRestrax}, 4F1, 4F2 \cite{SaclayFacilities} and SPINS \cite{SPINS}. However as neutron spectroscopy moves towards smaller sample sizes, the distance collimation can become comparable to or better than the mosaicity of standard graphite analysers. It has been shown that relying on distance collimation instead of mosaicity and the conventional parallel beam approximation can lead to better performing monochromators \cite{OptimalMono} so it would be natural if the same was true for analysers. We will show that this is indeed the case and additionally demonstrate the opportunity to analyse several energy bands simultaneously with a single analyser.
First we will describe the geometric effects in scattering from a single analyser slab, and then move to more advanced focusing and multiplexed setups. Finally we show how our ideas are verified by both experiments and simulations.

\section{Instrument and simulations}\label{sec:IaS}
The concept discussed in this article was developed for the CAMEA inverse time-of-flight spectrometer proposed for the European Spallation Source (ESS) \cite{ESSConcept}. Though the ideas are applicable to many crystal analyser spectrometer designs it will be discussed based on the 5 meV CAMEA analyser as this specific setup have been thoroughly investigated. The analyser will be placed at $L_{SA}=1.46$ m from the sample. It consists of 3*5 analyser crystals that are 1.0 cm wide, 5.0 cm long, and reflecting out of the horizontal scattering plane to several parallel $^3$He 1/2" (1.27 cm) linear position sensitive detector tubes  $L_{AD}=1.25$ m away. The settings are optimised for sample heights up to $h_{S}=1.0$ cm. All the work is done based on these settings unless stated otherwise. Simulations were performed using the McStas Monte Carlo ray-tracing package \cite{McStas1,McStas1.7}.

\section{Elements of the prismatic analyser concept}
The prismatic analyser uses a combination of distance collimation and an auto focusing effect from the analysers to achieve its results. We will here describe these effects before explaining the prismatic analyser itself.

\subsection{Distance collimation}
\begin{figure}
\begin{centering}
\includegraphics[width=0.45\textwidth]{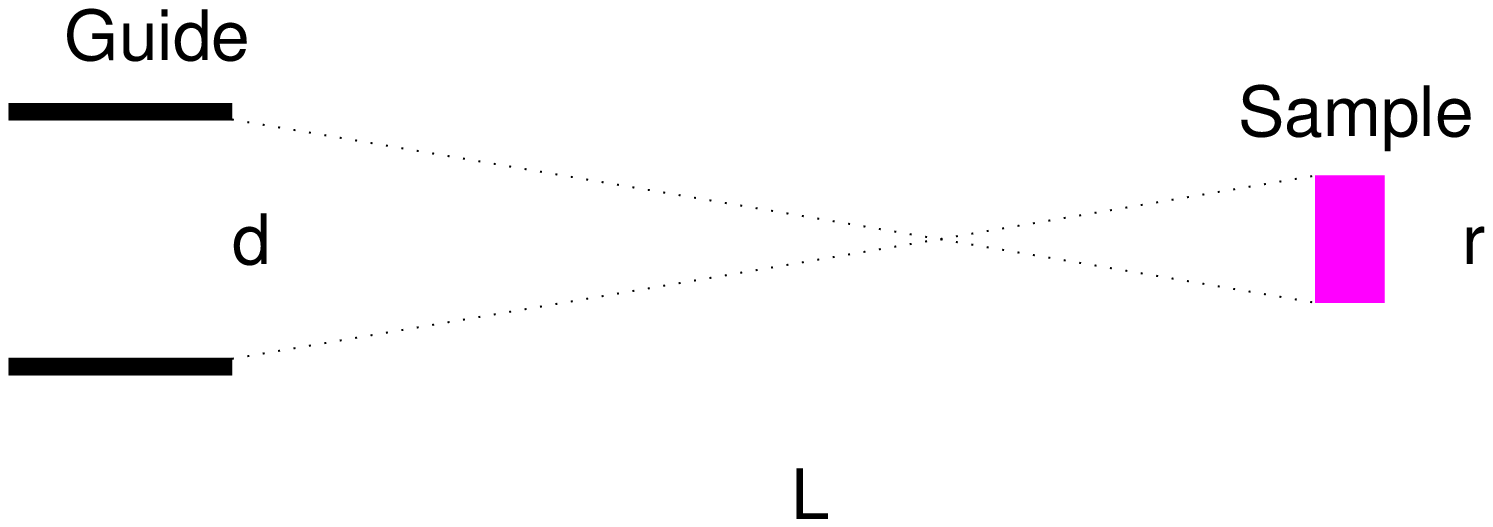}\quad \includegraphics[width=0.45\textwidth]{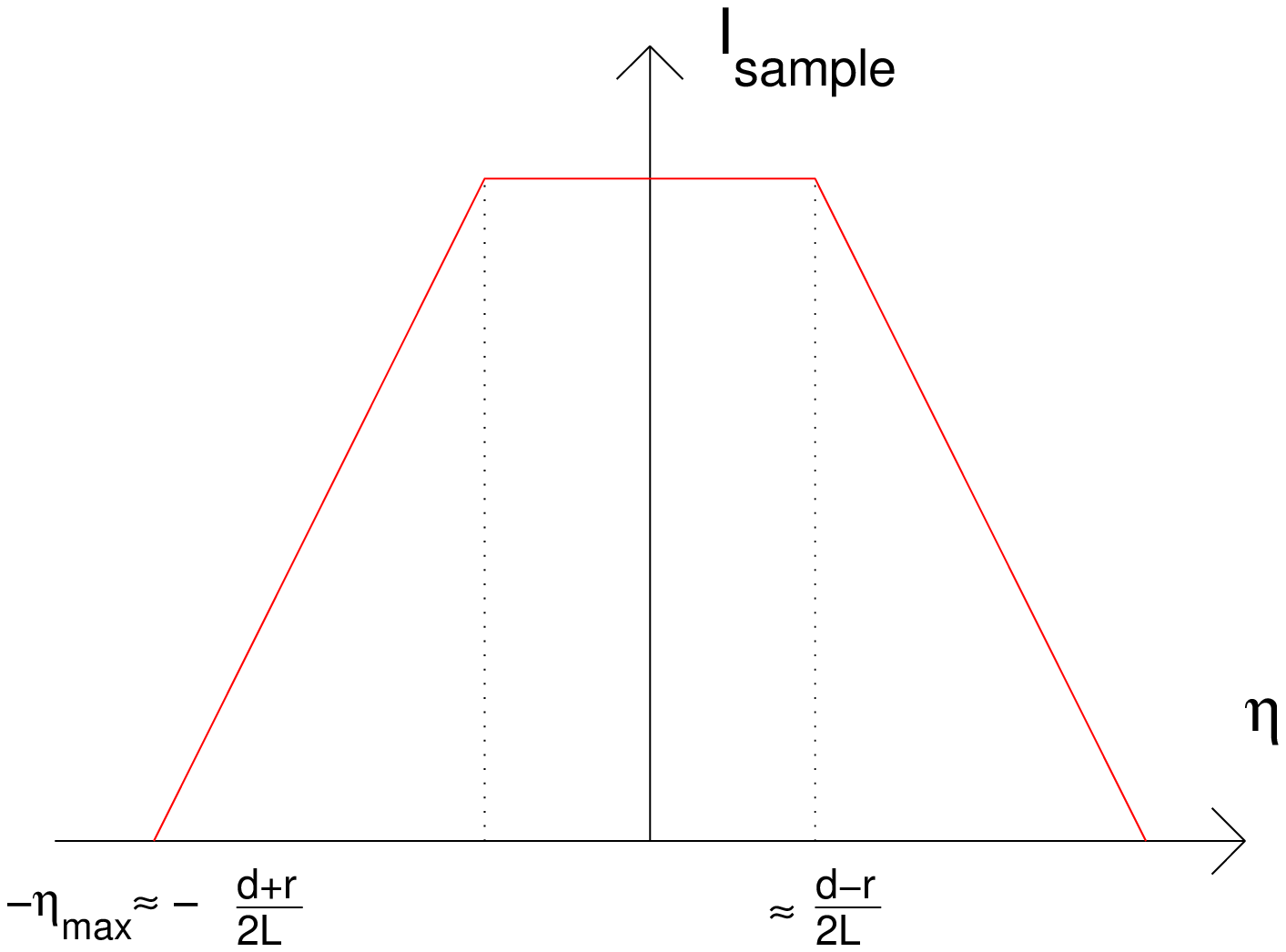}
\noindent\rule{0.9\textwidth}{0.4pt}\\
\includegraphics[width=0.45\textwidth]{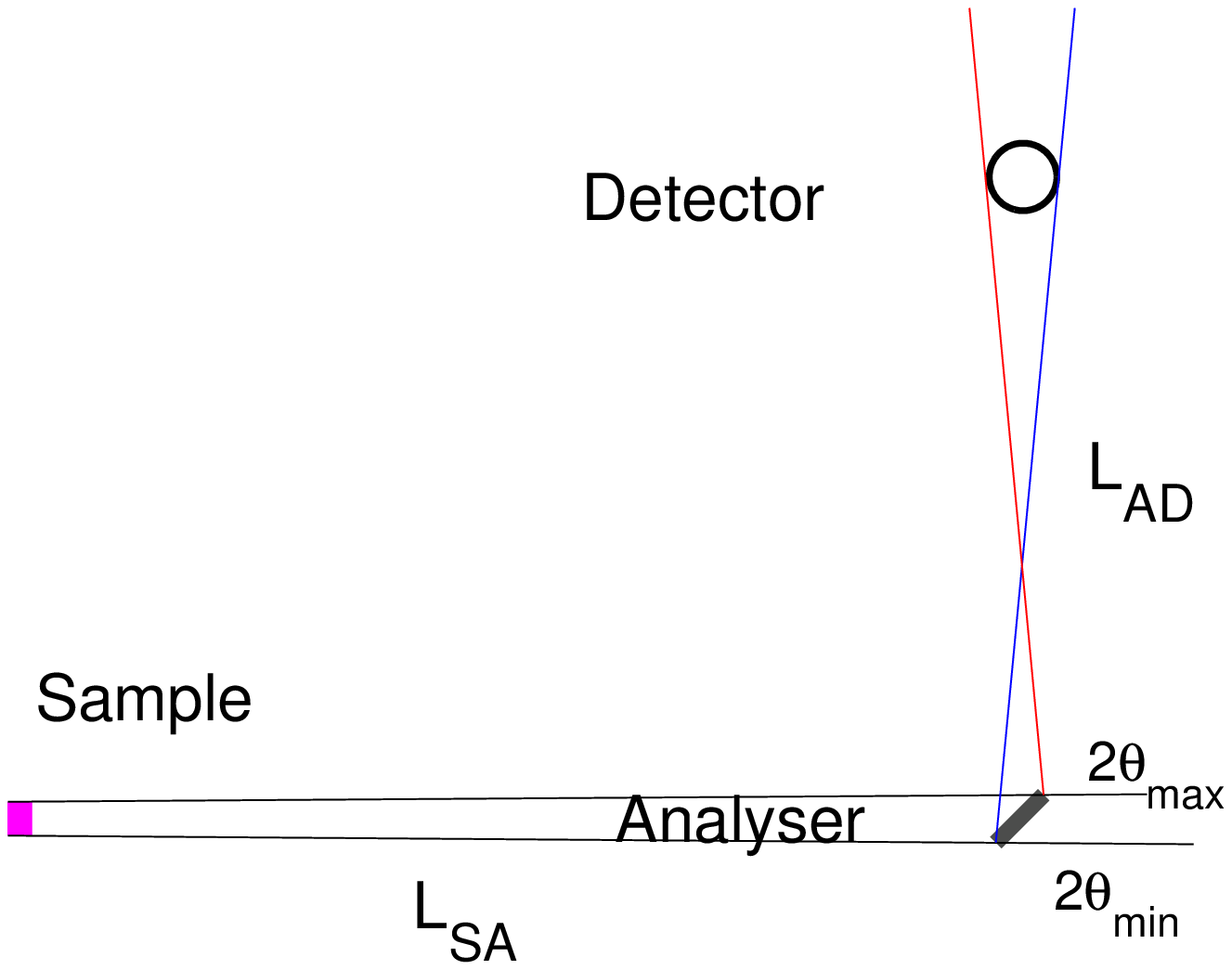}\quad \includegraphics[width=0.45\textwidth]{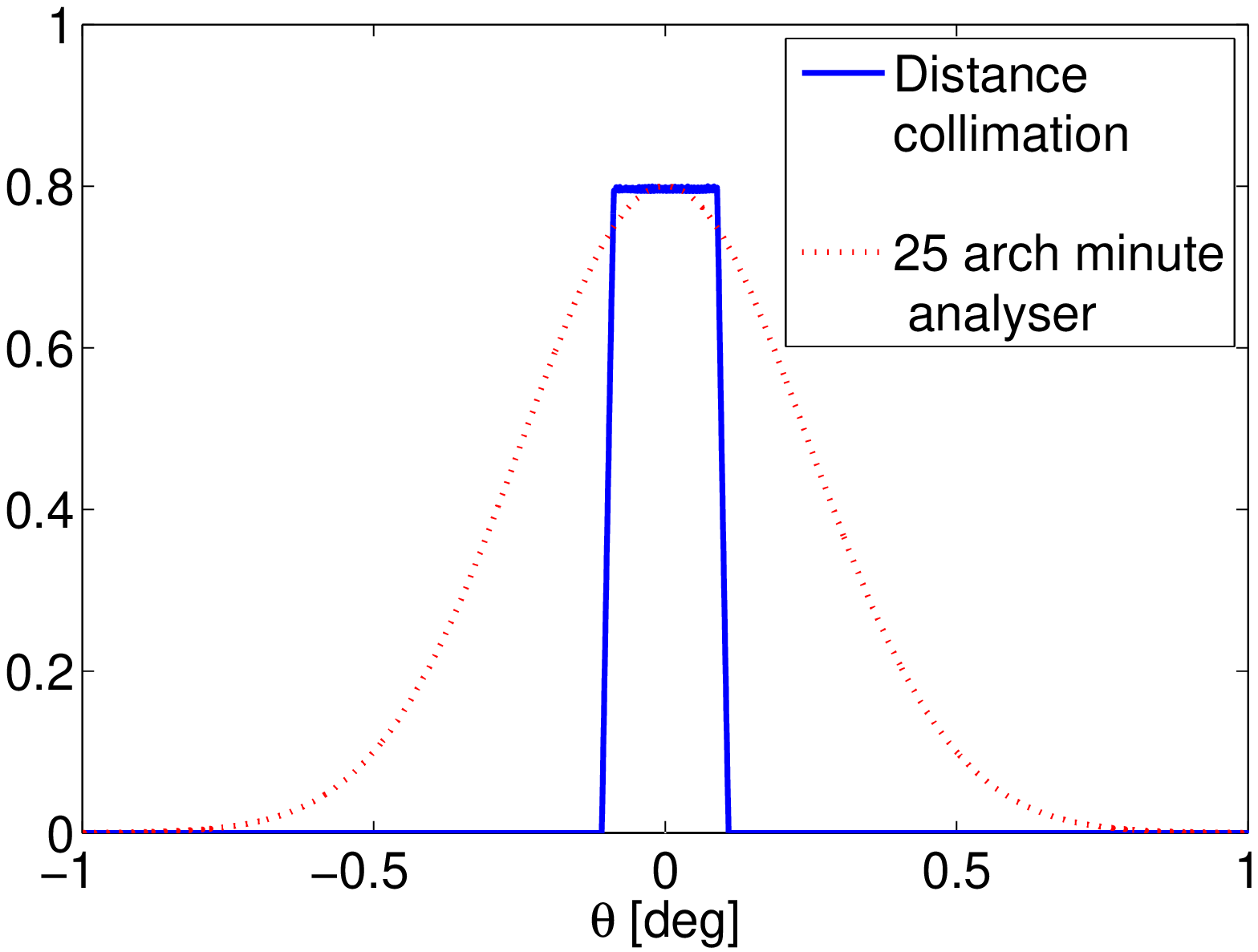}
\caption{Distance collimation. a) Geometrical constraints limits the possible paths from the guide to the sample. This will lead to a divergence distribution on the sample as shown in b), assuming uniform divergence and position distributions with no correlations at the end of the guide. c) The effect of distance collimation on analysers. Due to geometrical restrictions only polychromatic scattering with a Bragg angle between $2\theta_{min}$ and $2\theta_{max}$ can reach the detector independently of the analyser mosaicity. The corresponding rays that cross between sample and analyser have less extreme angles when sample and detector are roughly equal in size and $L_{SA}>L_{AD}$. d) Comparison of the resolution from distance collimation (numerically calculated) and 25' analyser mosaicity for the reference 5 meV CAMEA analyser.} 
\label{fig:DistCol}
\end{centering}
\end{figure}

Distance collimation is used in neutron instrumentation as a supplement to collimators to achieve a well collimated beam \cite{AdaptiveTripAx,RitaMonoim,MagOrdSpinW}. If two parts of an instruments (for example the guide end and the sample) have a maximum size and a certain distance between them then the maximal divergence that can make it through the instrument is limited (see figure \ref{fig:DistCol} a-b). These geometrical constraints are called distance collimation. In the prismatic analyser case we consider a variation: We have correlated distance collimation contributions between sample and analyser and between analyser and detector. We therefore consider the maximum variation in Bragg angle that allows reflection from somewhere on the sample via any spot on the analyser to somewhere on the detector (see figure \ref{fig:DistCol} c). For our reference setup this leads to a distance collimation (shown in figure \ref{fig:DistCol} d) of the order 12' FWHM and thus dominates the mosaicities of most graphite analysers. This makes it possible to relax the mosaicity further without any change in energy resolution.

\subsection{The auto-focus effect}
A monochromatic neutron beam will be focused at a certain distance by a single analyser slab. This "auto-focus" is illustrated in figure \ref{fig:principle}. Panel a) illustrates how a perfect monochromatic beam is reflected and focused by an analyser with a coarse mosaicity. Simulations on narrow energy bands c) - e) confirm the effect by a clear narrowing of the reflected beam at 80-100 cm from the analyser. The exact focusing spot will move further away (and be more focused) for smaller sample sizes so it is not possible to place the detectors in an exact focusing position for a general sample. However it is possible to construct the system so the auto focus will be close to the detectors for a wide range of sample sizes. 

\subsection{A single prismatic analyser}
\begin{figure}
\begin{centering}
\includegraphics[width=0.45\textwidth]{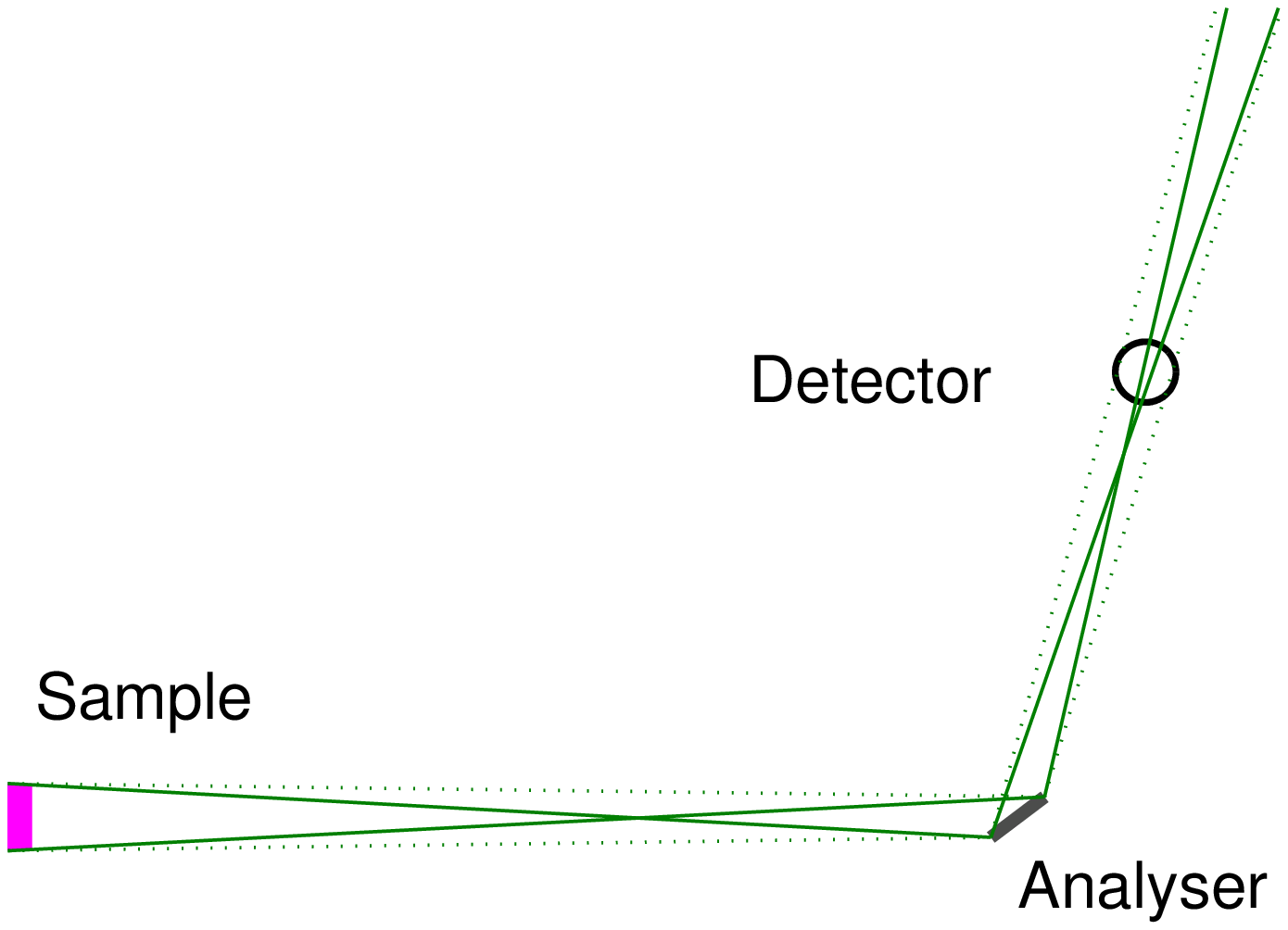}\quad \includegraphics[width=0.45\textwidth]{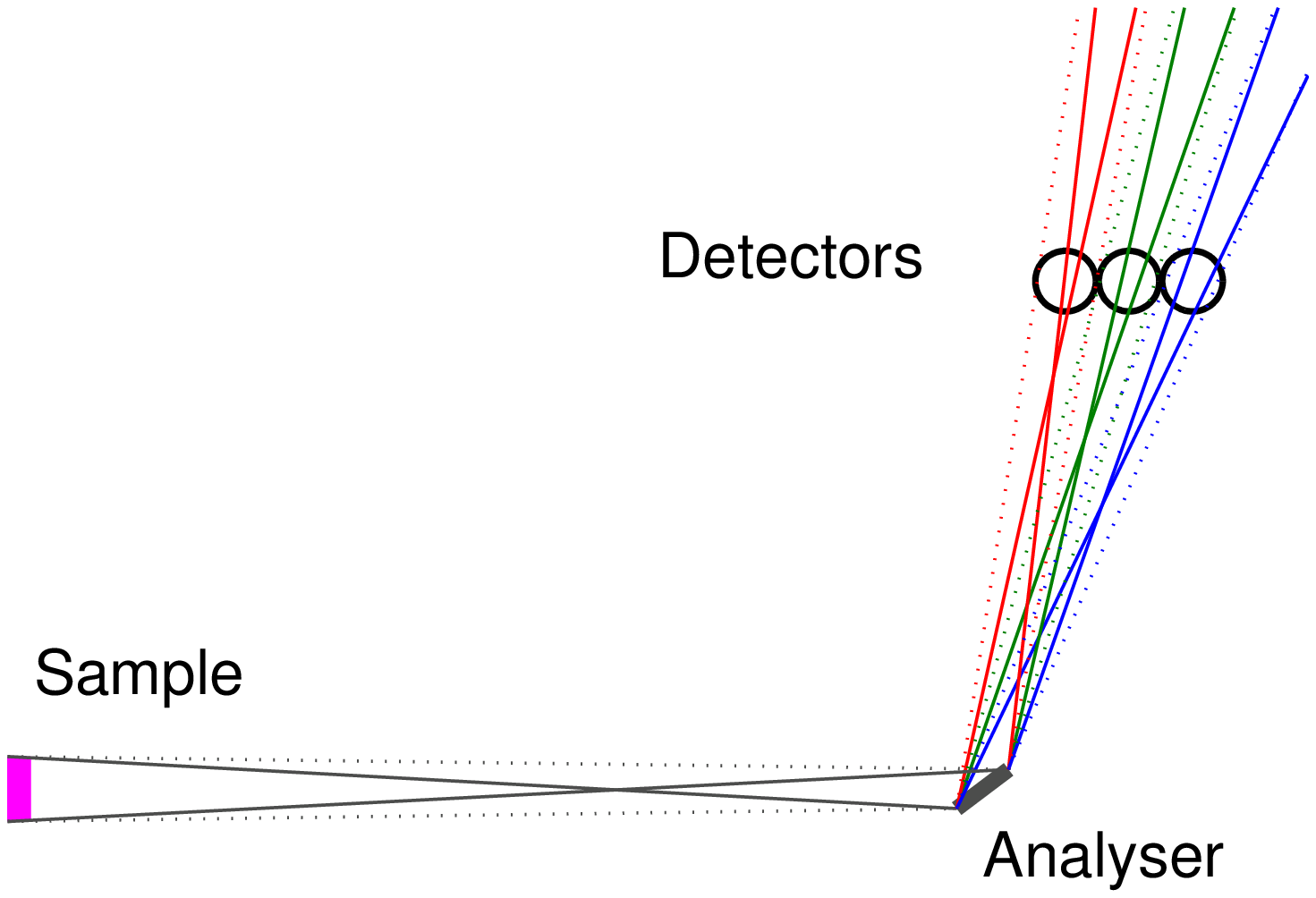}\\
\includegraphics[width=0.99\textwidth]{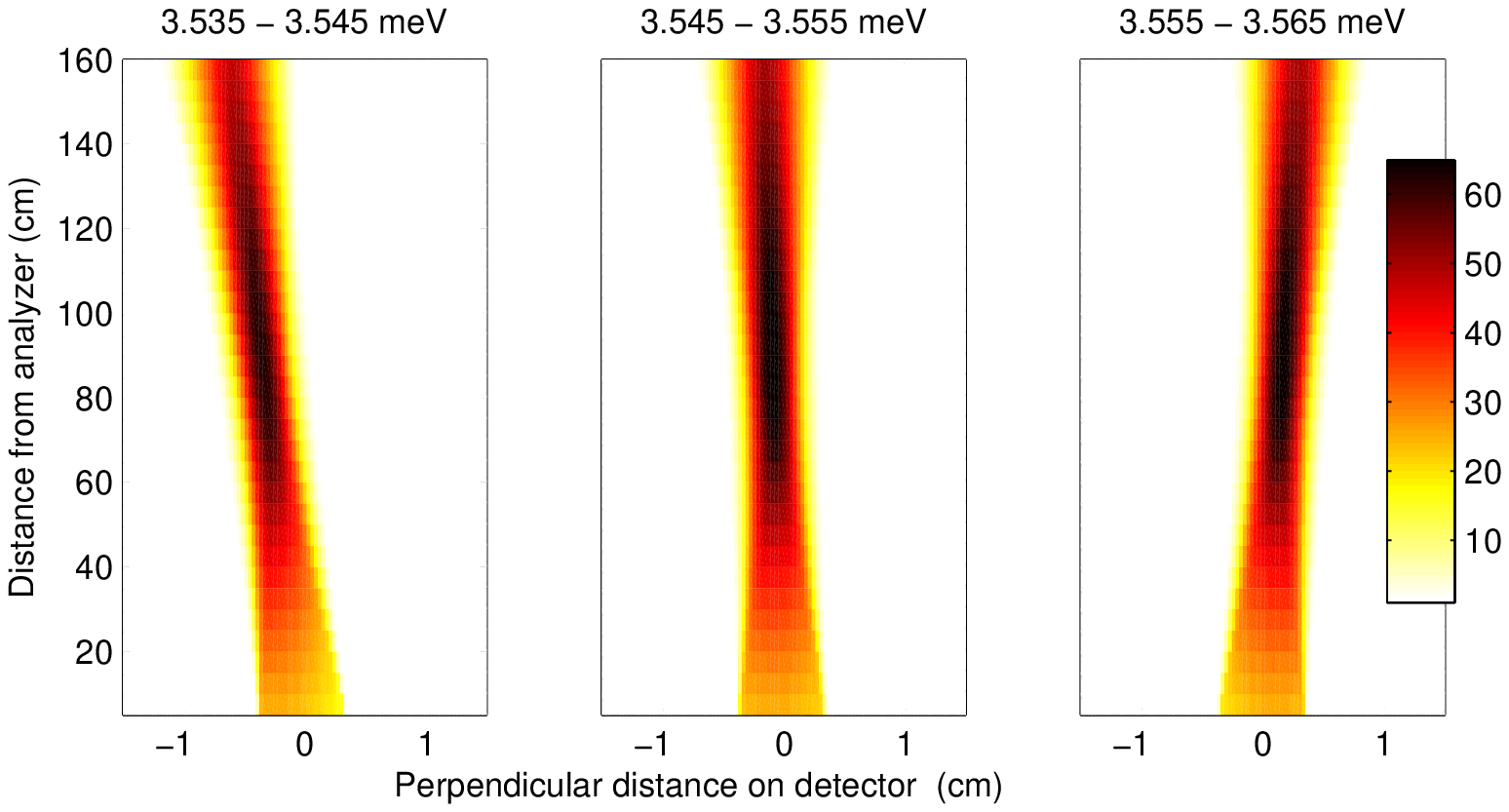}
\caption{a): Reflection of a monochromatic beam from a single analyser crystal focusing at a certain distance. The solid and dashed lines represent the limits of the scattered rays. The outer gives the width of the beam while the inner illustrates how focused the beam will be. b) Reflection of 3 specific energies from a single analyser crystal. Each energy is illustrated as in a) and reflected in a specific angle given by Braggs Law. The large difference in focusing distance is due to the exaggerated sample size, analyser size, and angular separation. c-e) McStas simulation of the beam profile as a function of distance from the analyser of 3 adjacent energy bands, 3.54 meV, 3.55 meV, and 3.56 meV from a single reflecting analyser piece.}
\label{fig:principle}
\end{centering}
\end{figure}

If distance collimation is the dominant part of the energy resolution the analyser crystal will reflect a wider energy band than measured by the detector. The remaining band will be reflected at slightly different angles as described by Bragg's Law, and thus miss the detector. E.g. the reference 5 meV analyser has a mosaicity of 60'=$1^\circ$ so the spread in scattering angle is $2^{\circ}$ (FWHM) and the real space FWHM of the reflected energy band at the detector position is 4.4 cm, substantially larger than the 1.27 cm width of the detector tube. However, due to the distance collimation each specific energy will be reflected into a much smaller angular band. Figure \ref{fig:principle} shows how 3 different monochromatic beams are reflected from the same analyser and recorded by 3 different detectors (b). McStas simulations of 3 narrow adjacent energy bands are shown in c) - e). Though there is some overlap it is clear that the energy affects the direction of the reflected beam. If a sufficient number of detectors are installed the entire flux reflected from the analyser will be recorded. In addition, the distance collimation provides an accurate determination of the Bragg angle. This provides a better resolution than most mosaicity limited spectrometers together with comparable total count rates from the same analyser.\\ 

\subsection{Simulated performance of the prismatic analyser}
\begin{figure}
\begin{centering}
\includegraphics[width=0.58\textwidth]{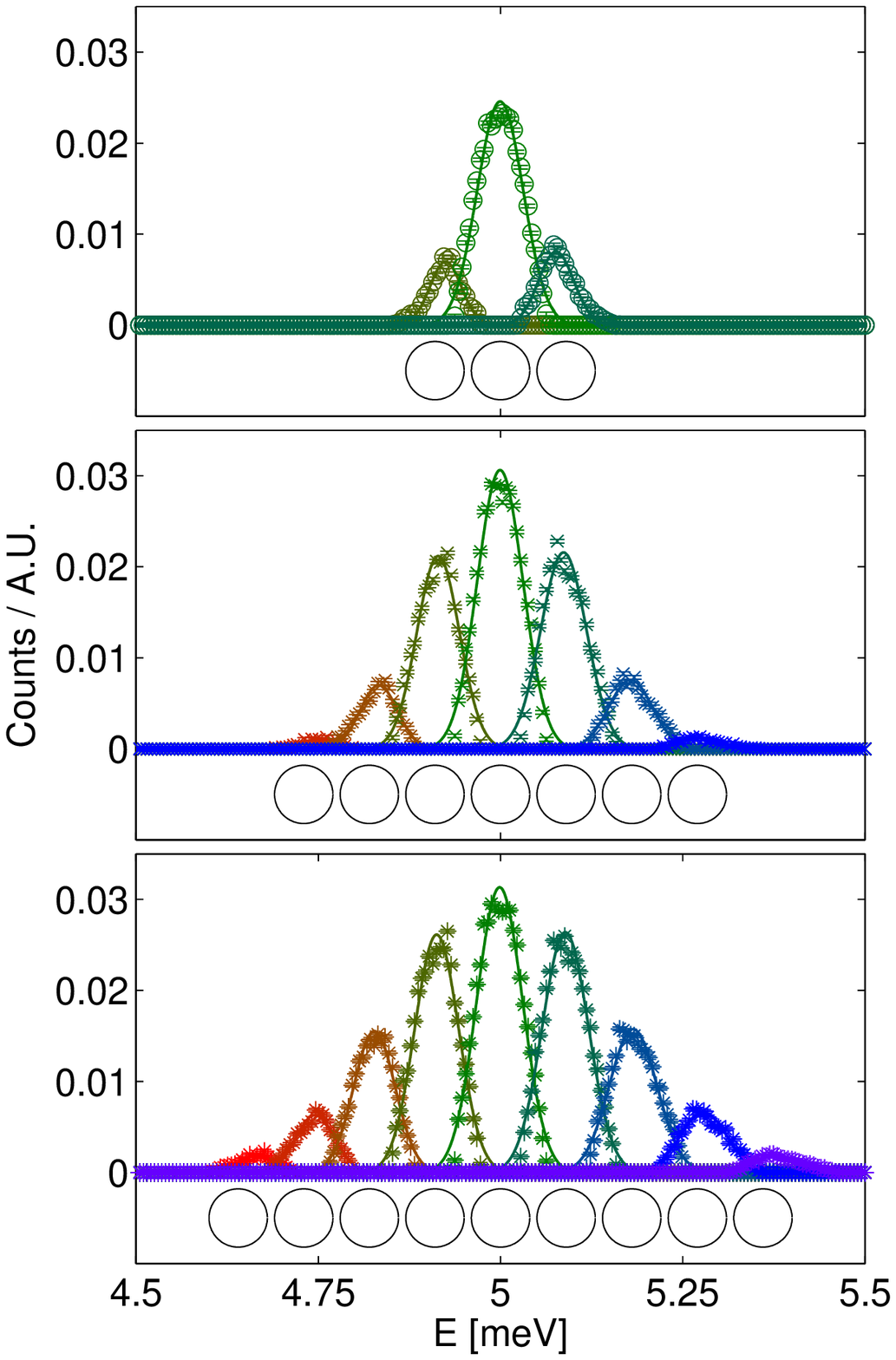}
\includegraphics[width=0.41\textwidth]{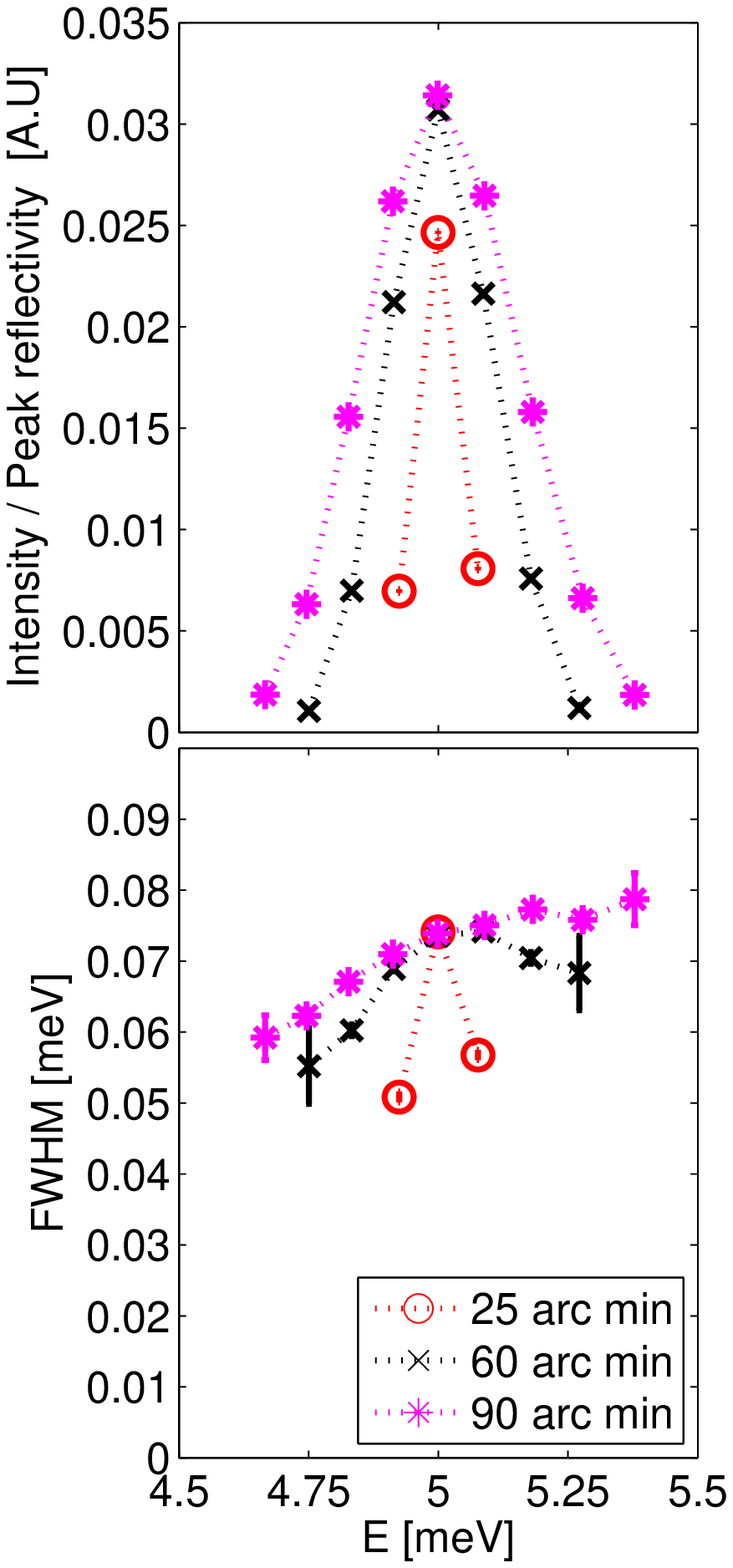}
\caption{Simulated recordings of several energies from a single analyser illuminated with a white beam. Each peak in a) - c) represents the counts in a single detector tube as function of E$_i$ (The detector tubes are represented with circles below the data). The mosaicity of the analyser is 25' for a), 60' for b), and 90' for c). d) Shows the corresponding intensities before correcting for peak reflectivity, and e) displays the resolution of the detectors for the 3 mosaicity values.}
\label{fig:MultitubeSim}
\end{centering}
\end{figure}

Figure \ref{fig:MultitubeSim} shows simulations of intensity and resolution for different mosaicities. We see that coarser analysers allow detection of more energies without affecting the resolution and will even detect more neutrons in the central detector. However coarser graphite will in practice lower the peak reflectivity counteracting this gain. Intensities should be scaled to reflectivity values. However peak reflectivity depends on analyser thickness, manufacturing process and the reflected energy \cite{Reflect1,GraphiteRef2} and can only be applied once these things are determined. The resolution broadens with higher energy as expected \cite{tripAxBook}. The better resolutions at the outer detectors, especially at the 25' analyser, are due to the analyser illuminating one part of the detector tubes more than the other. Thereby the effective detector width decreases which improves the distance collimation.\\
The outermost detectors will have much smaller intensities than the central and can be omitted to get comparable statistics and signal-to-noise in the different channels. Even though coarser mosaicity will lower the peak reflectivity, it will increase the total count rate provided there are enough detectors. The same effect can be achieved replacing the detector tubes with one position sensitive detector. However, in this work we will concentrate on a detector setup with thin tubes. 

\section{Advanced instrument designs}\label{sec:advancedGeometries}
In addition to the improved performance offered by the prismatic analyser there are several other techniques to improve the performance of triple axis-type spectrometers, such as focusing and multiplexing. This new concept will be most useful if combined with these techniques. 

\subsection{Focusing analysers}
\begin{figure}
\begin{centering}
\includegraphics[width=0.45\textwidth]{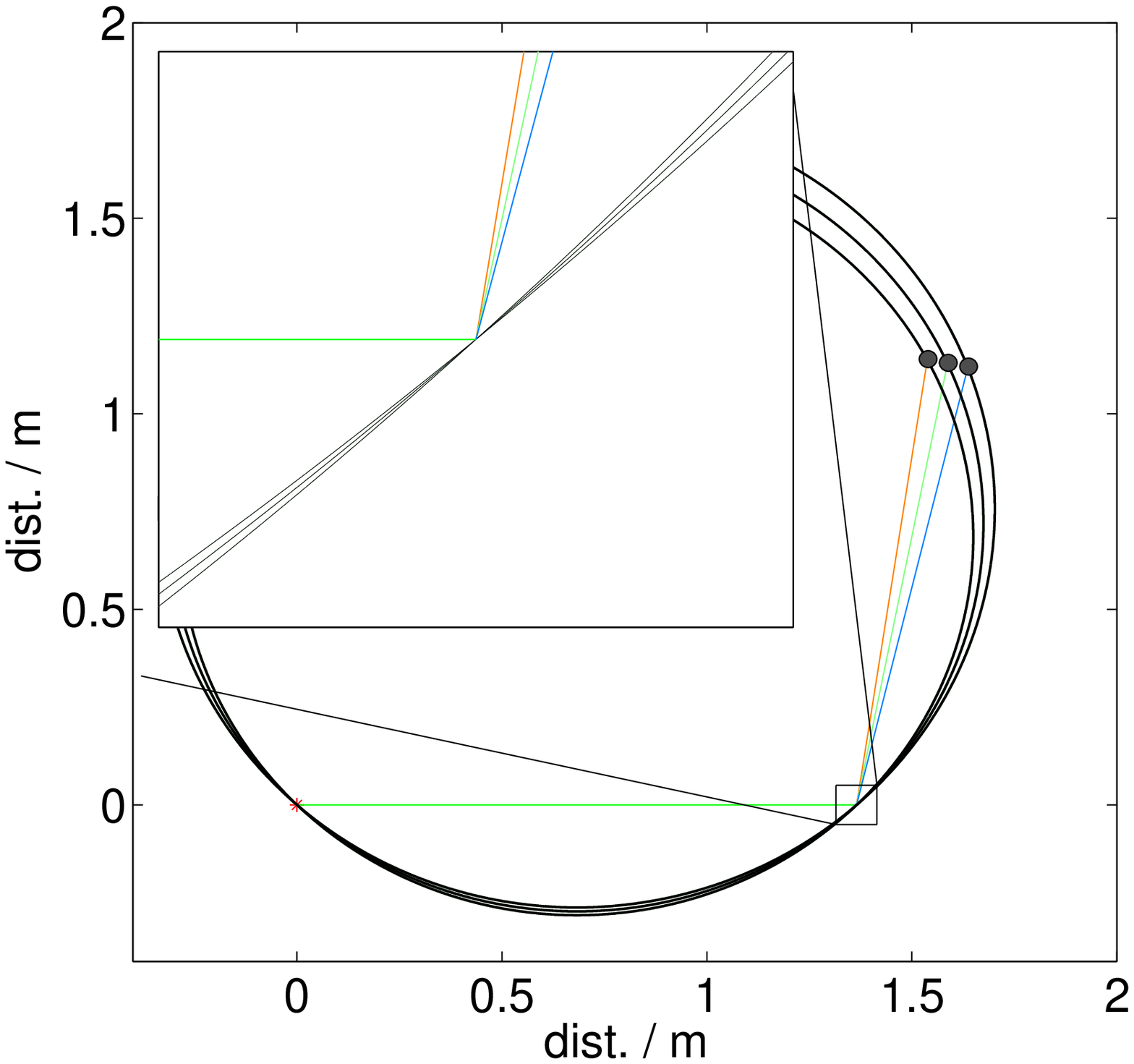}\quad \includegraphics[width=0.45\textwidth]{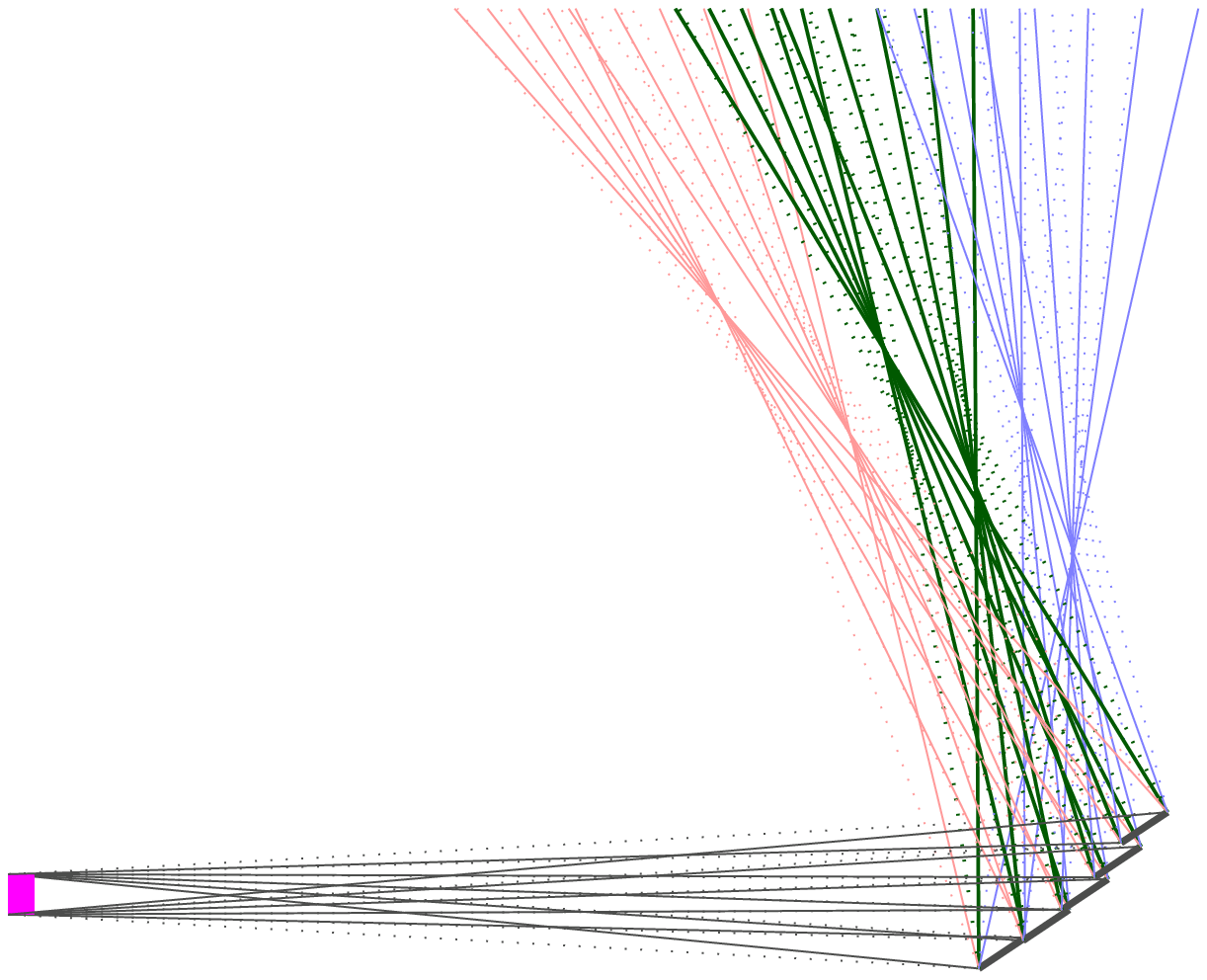}
\caption{a) Illustration of the optimal Rowland Circles for 3 parallel 1/2" detector tubes. At the analyser positions the circles are very close together, making it possible to get almost perfect focusing for all 3 detectors from one focusing analyser. b) The principle shown in figure \ref{fig:principle} for 5 analysers arranged in a Rowland geometry, the 3 energies represented by the 3 different colors are separated at the focusing distance.}
\label{fig:focusing}
\end{centering}
\end{figure}

An important component in distance collimation is the limited analyser width that unfortunately limits the covered solid angle. However like in conventional analyser spectrometers, this can be countered by arranging the analysers in a focusing Rowland Geometry \cite{HabichtRowland}. The Rowland Geometry is very robust to small perturbations in energy so the outer detectors will be in almost perfect focusing condition when the analyser is focused on the central detector. Figure \ref{fig:focusing} a) displays the optimal Rowland circles for reflecting 3 different energies towards 3 different detectors from the same analyser position. At the analyser the distance between the circles are smaller than the width of the analyser crystals so the focusing is almost perfect for all detectors independent of which of the circles is chosen. \ref{fig:focusing} b) shows the schematics of how 3 different energies are reflected and how they can be separated at the detector position. The crystals are chosen to be so close that no gap is seen from the sample, but that does lead to a small overlap between crystals seen from the detector. Simulations have shown that this shadow effect is negligible and for practical purposes the finite width of crystals and mounting will anyway force the analysers further apart. \\
By focusing it is possible to increase the solid angle coverage and thus improve the recorded flux just like with a conventional analyser setup. To confirm that it does indeed work we performed a full simulation with 5 analyser blades in a focusing geometry. This provided a factor 4.6 in flux gain without sacrificing energy resolution (data not shown).

\subsection{Multiplexing}
Multiplexing spectrometers have become increasingly popular with varying layouts like RITA II \cite{RitaExplanation,Realizing_RITA}, IMPS \cite{IMPS}, and Flatcone \cite{Kempa20061080}. A challenge when combining multiplexing with prismatic analysers is that many multiplexing instruments have several detectors close together measuring reflections from different analysers. There might thus not be sufficient space for the optimal number of detectors. However by choosing slightly sub-optimal settings it is still possible to combine the two techniques.\\ 
For example the proposed ESS CAMEA will have a multiplexing setup with a very large analyser coverage. It has 10 concentric rings of analysers reflecting 10 different prismatic energy bands, as seen in fig. \ref{fig:principle}, to position sensitive detectors below the scattering plane\cite{ESSCAMEA}. While this extreme case of multiplexing could be combined with any number of detectors pr analyser, a detector number above 3 would force the innermost analysers further apart than optimal and impose severe extra costs. In contrast 3 detectors can be included without any drawbacks.

\section{Experimental verification}
\begin{figure}
\begin{centering}
\includegraphics[width=0.70\textwidth]{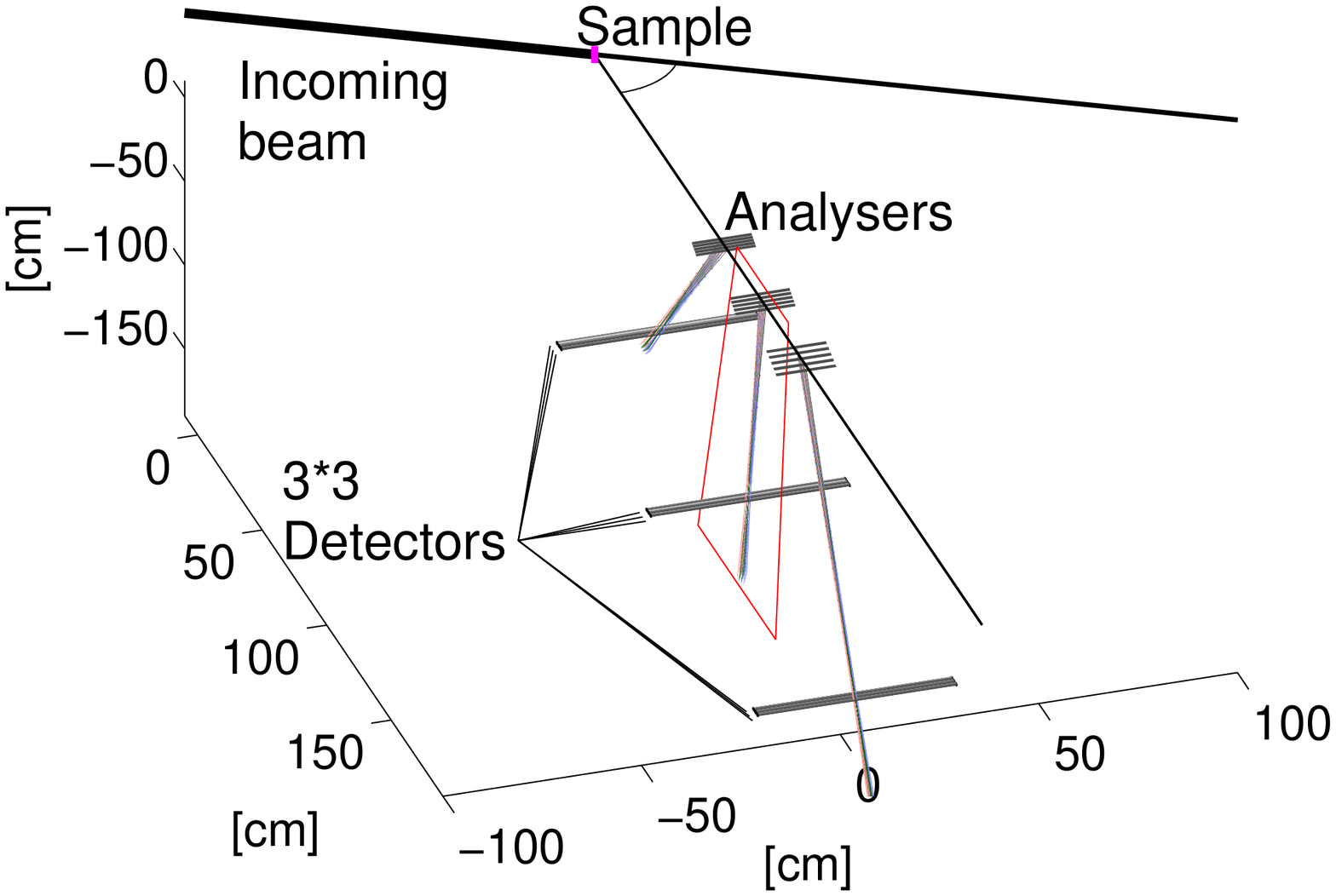}\includegraphics[width=0.25\textwidth]{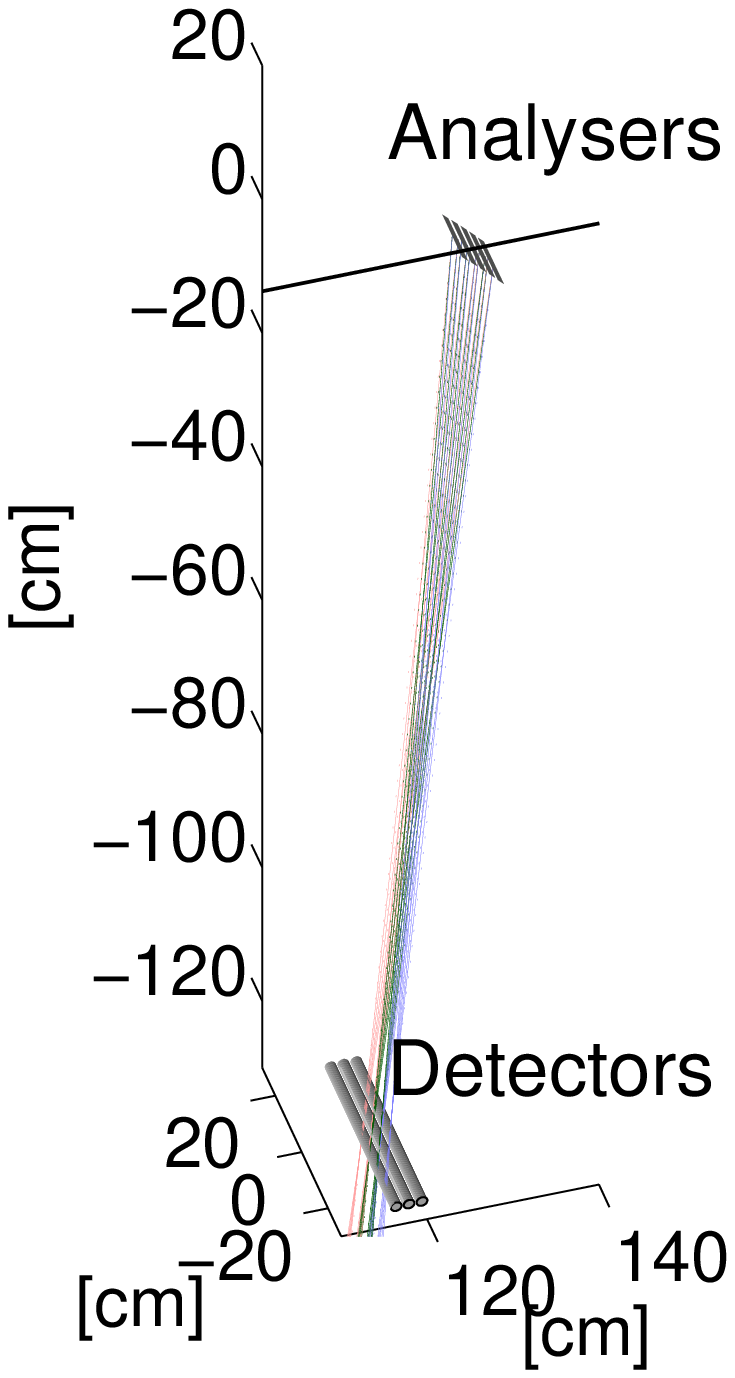}\\
\includegraphics[width=0.60\textwidth]{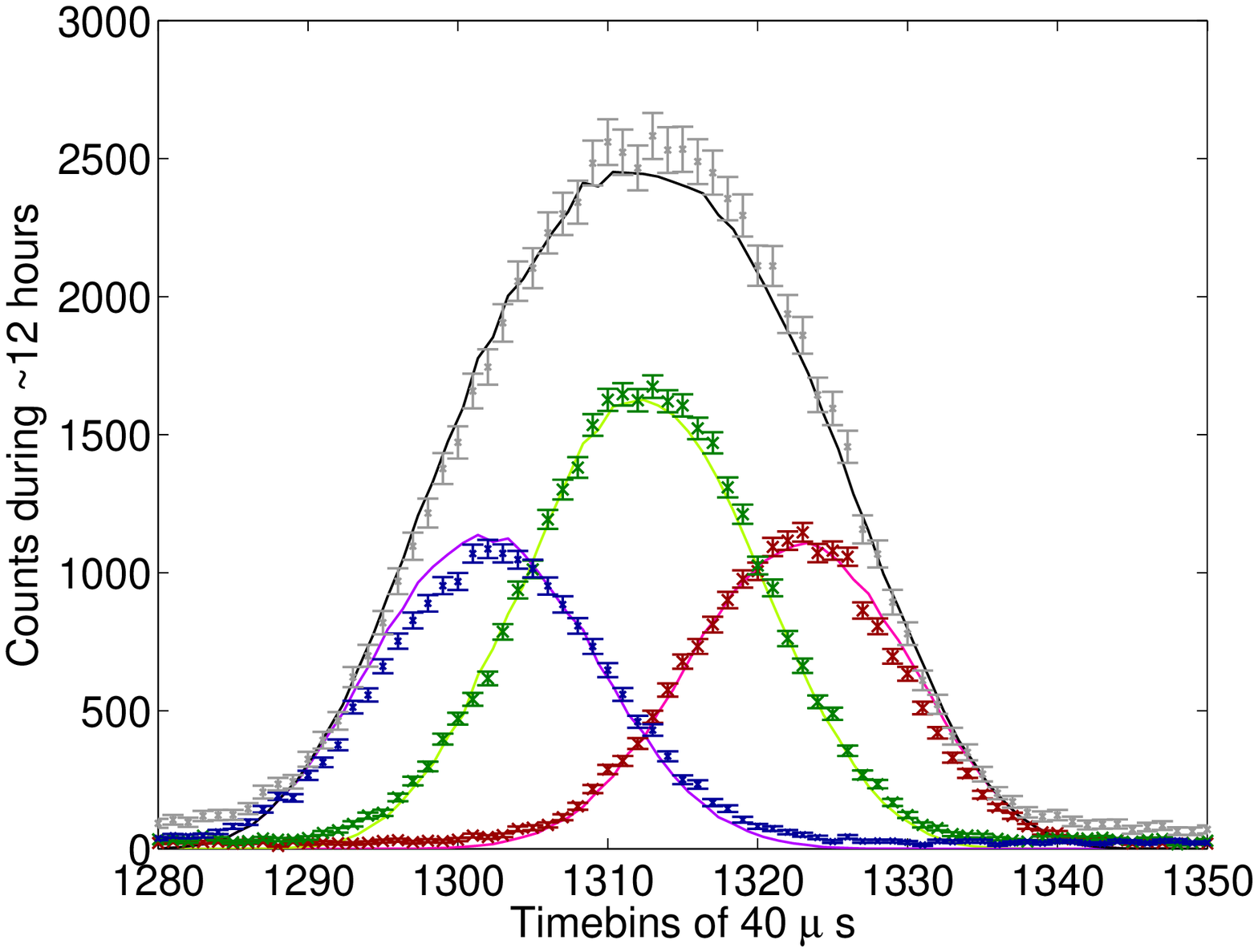}
\caption{a): Sketch of the experimental setup of the CAMEA prototype. The analyser-detector setup in the red box is shown in b). Data from this single analyser-detector setup is below. c) Time distribution of neutrons scattered by a 2.2 cm tall Vanadium sample and detected in each of the 3 detectors recording data from the 5 meV analyser. Measured data are given by data points and simulated data by lines. The coloured peaks show the result of using the prismatic analyser while the grey shows the corresponding signal from adding all 3 detectors together and relying on 30' mosaicity for energy resolution. The simulated intensities have been rescaled by one common factor in order to compare the line shapes. The data is displayed in raw time bins. Each time bin of 40 $\mu$s corresponds to $\sim 10$ $ \mu eV$. The technique has also been tested at other distances and mosaicities and found to work there as well. 
	}
\label{fig:Prototype}
\end{centering}
\end{figure}

A prototype of the ESS CAMEA prismatic analyser was built at Ris\o ~campus of DTU and installed on the MARS backscattering spectrometer at PSI \cite{MarsReport} in 2012.  MARS is an inverse time-of-flight instrument with a flight path between the master chopper and sample of 38.47 m. 
The prototype consists of 3 vertically focusing analysers behind each other. Each consists of five 15 cm wide, 1 cm tall analysers in Rowland geometry and scatters the neutrons out of the plane to 3 linear position-sensitive detectors. The distance between sample and analyser is 1.2 m and between analyser and detectors 1.0 m. The analysers are centred around a $2\theta$ value of $60^\circ$. Due to spatial restrictions in the prototype the test was not done at the exact settings proposed for the final instrument. A more thorough description of the prototype and its testing will be reported elsewhere. \\
Figure \ref{fig:Prototype} shows data obtained from the prototype experiment. In b) we used a Vanadium sample to ensure incoherent elastic scattering and recorded the energy separation expressed as neutron time-of-flight in the 3 detectors. The simulations were done at the same settings but the intensities were rescaled with one common factor to account for imprecise descriptions of source brilliance, sample volume, and analyser peak reflectivity. The data is displayed in the raw time bins in order not to impose any data treatment assumptions. The data confirms that it is indeed possible to separate several energy bands and obtain the good resolution promised by the simulations from a focusing prismatic analyser in a multiplexing inverse time-of-flight spectrometer.

\section{Comparison to a conventional spectrometer}
Examples of simulated gain factors for prismatic analysers are shown in table \ref{tab:Comparison}. The comparison is done for a vertical Rowland geometry as described in \ref{sec:IaS}, and for a reference using the same geometry but 25' analysers and a single detector taking up the same space as the full detector setup of the prismatic analyser. So it is 3.95 cm wide (including the 1 mm spacing between the detector tubes) when compared to 3 detectors and 6.65 cm wide when compared to 5 detectors. The Rowland setup already gives the reference a gain of 1.7 in flux and 1.7 in resolution reduction compared to a single 5 cm analyser.	The resolution reduction is understood as $\sigma_{r}/\sigma_{new}$ where $\sigma_{new}$ is determined from the central 5 meV detector and $\sigma_{r}$ is the resolution of the reference setup. The intensity gain is defined as $R_{\eta}/ (I_r \cdot R_{25'}) \cdot \sum_n I_n   $ where $I_n$ is the intensity of the n'th tube looking at a given analyser and $I_r$ is the intensity on the reference detector.  $R_{\eta}$ is the peak reflectivity of the analyser with mosaicity $\eta$. For this comparison we used typical peak reflectivity values of 0.8, 0.7, and 0.6 for 25', 60' and 90' respectively.\\
The results demonstrates that it is possible to improve the resolution a factor of 2 while at the same time doubling the intensity, corresponding to a total gain factor of 4 compared to a traditional mosaicity driven analysers. The gain factors can be increased slightly by using position sensitive detectors.  
\begin{table}
	\centering
		\begin{tabular}{|c |c||c|c|}
		\hline
			 Analyser mosaicity&No. of detector tubes & Resolution reduction & Intensity Gain\\
			\hline
			25' &3 &  2.0& 0.9\\
			\hline
			60'  & 3 &  2.0 & 1.4\\
			60' & 5 &  2.0 & 1.9\\
			\hline
			90' & 3 & 2.0 & 1.4\\
			90' & 5 & 2.0 & 2.0\\
			90' & 7 & 2.0 & 2.3\\
			\hline
		\end{tabular}\\
	\caption{Gain factors for different prismatic analyser layouts obtained by simulations.}
	\label{tab:Comparison}
\end{table} The 0.9 in flux gain factor for 3 detectors and 25" mosaicity is due to the difference between the 3 round detectors with less efficient edges and spacing between them and the single big square detector of the reference model.\\
As discussed there is only space for 3 detector tubes per analyser on CAMEA and thus the 3 tubes 60' mosaicity has been chosen for this design. This reduces the resolution a factor 2.0 and increases the flux a factor 1.9 for the ESS version when compared to a traditional Rowland geometry with the same analyser and detector area. Compared to a flat analyser slab one gains a factor of 3.3 in resolution reduction and 3.2 in flux or a total gain factor of 10.6.\\

\section{Conclusion}
Cystal analyser spectrometers designed for small samples have a better distance collimation than standard triple axis spectrometers. Instead of reducing the distance collimation or accept lower count rates the geometric constraints can be used as a benefit by installing several detectors that record different energies from the same analyser. If the mosaicity is relaxed, this can produce better resolution and higher total count rates than achievable by installing fine mosaicity analyser crystals at the spectrometer or by using Soller collimators. The method is proven by both measurements and simulations to work together with analysers arranged in Rowland geometries and multiplexing setups.\\
We have further exemplified that a 60' mosaicity setup with 3 detector channels can lead to an resolution improvement of a factor 2.0 together with a flux increase of a factor of up to 1.9 compared to a conventional 25' mosaicity single detector Rowland setup. Even bigger gain factors of 3.3 in resolution reduction and 3.2 in flux can be achieved when compared to a flat analyser slab.
 
\section*{Acknowledgements}
This project was funded by the Danish and Swiss in-kind contributions to ESS design and update phase. We thank Astrid Schneidewind for insightful comments.


\bibliographystyle{h-physrev.bst}
\bibliography{MultiWave}





\end{document}